\documentclass[a4paper,11pt]{article}

\usepackage{jheppub}

\usepackage{latexsym}
\usepackage{epsfig}
\usepackage{graphicx}
\usepackage[english]{babel}
\usepackage{amssymb}

\title{Rapidity and momentum transfer distributions of coherent $J/\psi$ photoproduction in 
ultraperipheral pPb collisions at the LHC}

\author{V. Guzey}
\author{and M. Zhalov}

\affiliation{National Research Center ``Kurchatov Institute'',
Petersburg Nuclear Physics Institute (PNPI), Gatchina, 188300, Russia}

\emailAdd{vguzey@pnpi.spb.ru}
\emailAdd{zhalov@pnpi.spb.ru}


\abstract{
Based on accurate calculations of the flux of equivalent photons of the proton 
and heavy nuclei and
the pQCD framework for the gluon distribution in the proton and nuclei, we
analyze the rapidity and momentum transfer distributions of coherent $J/\psi$ photoproduction in 
ultraperipheral proton--Pb collisions at the LHC. 
We demonstrate that unlike the case of proton--proton 
UPCs marred by  
certain 
theoretical uncertainties and experimental limitations, 
after a cut excluding the region of small momentum transfers, ultraperipheral proton--Pb 
collisions offer a clean way to study the gluon distribution in the proton down to 
$x \approx 10^{-5}$.
Our analysis of the momentum transfer distributions shows that
an interplay of  $J/\psi$ production by low-energy photons on the nucleus
and by high-energy photons on the proton in proton--Pb UPCs 
can result in some excess of events at small $p_t$ in a definite region of the rapidity $y$.
}

\begin{document} 
\maketitle
\flushbottom

\section{Introduction}
\label{sec:into_gz}

In this paper, 
we discuss sources of the model dependence of
the analysis of $J/\psi$ photoproduction on the proton
in proton--proton ($pp$)
ultraperipheral collisions (UPCs) and show that the study of this process in 
proton--nucleus ($pA$) UPCs
 is free from these shortcomings.
Part of our results presents an update of the earlier prediction~\cite{Frankfurt:2006tp} 
for coherent $J/\psi$ photoproduction in 
ultraperipheral proton--Pb collisions (for reviews of 
high-energy ultraperipheral collisions and proton--nucleus collisions at the LHC, see
Refs.~\cite{Baltz:2007kq} and \cite{Salgado:2011wc}, respectively).

The recent ALICE~\cite{alice1,alice2} and LHCb~\cite{lhcb} measurements of exclusive
$J/\psi$ photoproduction in  
Pb-Pb and proton--proton UPCs at the LHC
confirmed the expectations~\cite{Baltz:2007kq} that UPCs are a very promising
way to study  the gluon distributions in nuclei  and the proton at small $x$.
In particular, 
the leading order pQCD analysis~\cite{Guzey:2013xba,Guzey:2013qza}  of the cross section 
of exclusive $J/\psi$ photoproduction in PbPb UPCs~\cite{alice1,alice2}  
allowed one---for the first time---to establish the evidence of 
the large nuclear gluon shadowing 
at $x \approx 10^{-3}$.

The LHCb measurements of the yield of $J/\psi$ 
at forward rapidities ($2<y<4.5$) in proton--proton 
UPCs at 7 TeV~\cite{lhcb} resulted in the extension of
the small-$x$ region previously studied at HERA in photon--nucleon scattering 
down to $x=6 \times 10^{-6}$.
The analysis of the data 
confirmed the power law energy dependence
of the $\gamma p\to J/\psi p$ cross section 
($\sigma (W_{\gamma p})\propto  W^{\delta}_{\gamma p}$ with $\delta=0.92 \pm 0.15$~\cite{lhcb}) 
consistent with the previous HERA results and
did not reveal any evidence of new phenomena such as an
onset of the gluon saturation regime at small $x$.
However, this conclusion should be considered 
preliminary
because of the large experimental errors 
and certain
theoretical uncertainties in the data analysis.

There are two main problems with studies of exclusive $J/\psi$ photoproduction in
proton--proton UPCs at the LHC. First, for symmetric (same energy) collisions and
in the situation when both protons in the final state remain intact, 
it is not possible to select kinematics
allowing one to determine
which proton emitted the photon and which one served as a target 
(the LHC detectors have not been capable so far to detect these protons because of  
their very small transverse momenta). 
As a result, 
the cross section of $J/\psi$ production in proton--proton UPCs is given by the sum of two terms
of a comparable magnitude. 
Each term can be calculated within the Weizs\"{a}cker--Williams (WW) approximation
as a product of the photon flux 
emitted by one of the colliding 
participants and the cross section of $J/\psi$ photoproduction on its partner:
\begin{equation}
\frac {\sigma_{AB\to ABJ/\psi}(y)} {dy}
=N_{\gamma/A}(y)\sigma_{\gamma B\to J/\psi B}(y)+
N_{\gamma/B}(-y)\sigma_{\gamma A\to J/\psi A}(-y) \,.
\label{csupc}
\end{equation}
In eq.~(\ref{csupc}), $A$ and $B$ stand either for the proton or a nucleus;
$N_{\gamma/A(B)}(y)$ is the photon flux;
$y = \ln (2\omega/M_{J/\psi})=\ln[W^{2}_{\gamma p}/(2\gamma_{L}^{A(B)}m_{N}M_{J/\psi})]$
is the $J/\psi$ rapidity, 
where $\omega$ is the photon 
energy, $W_{\gamma p}$ is the $\gamma p$ center-of-mass energy, 
$M_{J/\psi}$ is the mass of $J/\psi$, $m_N$ is the nucleon mass, and  
$\gamma_{L}^A$ and $\gamma_{L}^B$ 
are the Lorentz factors corresponding to projectiles $A$ and $B$, respectively.
In the case of proton--proton UPCs, provided that 
the photon flux is evaluated with good accuracy,
the cross section of $J/\psi$ photoproduction on the proton can in principle be 
reliably extracted from eq.~(\ref{csupc})
only in two cases: (i) at $y=0$, where both contributions in eq.~(\ref{csupc}) 
are equal since the energies of photons
emitted by both protons are equal, and (ii) in the region, where one of
the contributions dominates. 

\begin{figure}[htb]
\centering
\epsfig{file=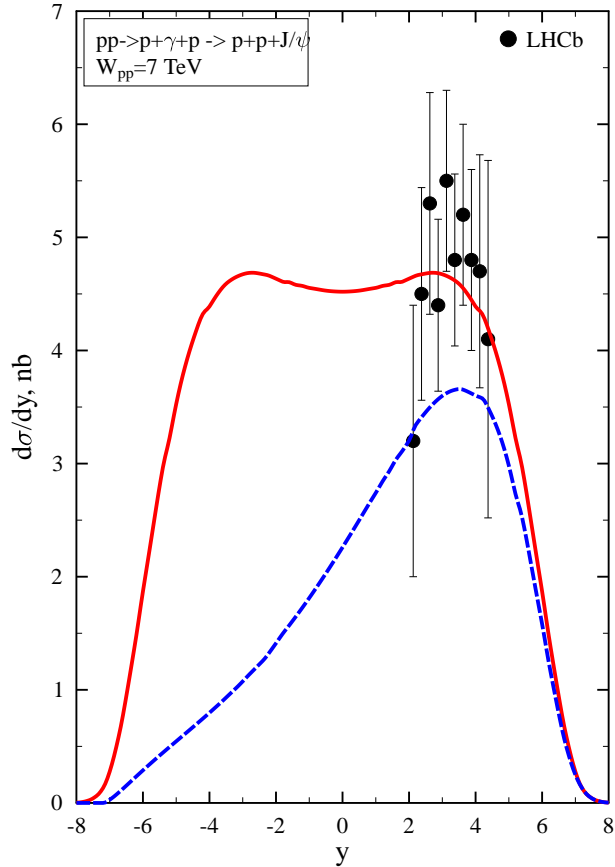,scale=0.5}
\vspace{-1 cm}
\caption{The rapidity distribution of $J/\psi$ photoproduction 
in proton--proton UPCs at $\sqrt{s_{NN}}=7$ TeV. The results of the LO pQCD calculation 
are given by the red solid [both terms in eq.~(\ref{csupc})] and the blue dashed
[first term in eq.~(\ref{csupc})] curves; the points and the corresponding
error bars are the LHCb data~\cite{lhcb}.}
\label{lhcb}
\end{figure}

However, one can demonstrate that the latter case is not realized at the LHC.
Using the leading order (LO) pQCD analysis of $J/\psi$ photoproduction on 
the proton~\cite{Guzey:2013qza}, we
calculate the rapidity distribution for exclusive $J/\psi$ photoproduction in 
proton--proton UPCs in the kinematics of the LHCb 
experiment~\cite{lhcb} 
(details of the calculation are discussed in section~\ref{sec:cs_gz}).
Our results are presented in figure~\ref{lhcb}: the red solid curve corresponds to
the sum of both terms in eq.~(\ref{csupc}); the blue dashed curve represents the contribution
of the first term in eq.~(\ref{csupc}).
One can clearly see from the figure that 
the two curves deviate from each other in the region of $y$ covered by the LHCb 
measurement: the term corresponding to photoproduction by low-energy photons
contributes at the level of 20\% in the rapidity range of $2<y<4.5$. This means that
the
dominance of either of the terms in eq.~(\ref{csupc}) is not realized in 
the LHCb kinematics.

The second problem with
proton--proton UPCs is related to the account for initial
and final state strong interaction between the colliding protons. 
In coherent photoproduction, strong interactions between 
colliding protons resulting in particle production should be suppressed, 
while elastic rescattering can still 
contribute~\footnote{Experimentally coherent $J/\psi$ events are selected by requiring 
only two leptons from the $J/\psi$ decay and otherwise an empty detector.}.
Some estimates of this suppression~\cite{Khoze:2002dc,Schafer:2007mm} 
predict the
suppression effect at the level of 20\% for $J/\psi$ photoproduction
at central rapidities in proton--proton UPCs at the LHC energies; 
the suppression increases 
with an increase of rapidity since higher photon energies require a
more significant contribution of small impact parameters. Besides, at 
very high photon energies,
one can expect an increasing role of photon emission in inelastic 
transitions~\cite{Bertulani:1987tz} 
 and even the breakdown of the WW approximation and, hence, invalidity of eq.~(\ref{csupc}).

We argue
that the study of $J/\psi$ photoproduction on the proton
is much more preferable in proton--nucleus UPCs than in proton--proton UPCs.
First, in the case of $pA$ UPCs, the collision is asymmetric and, hence, coherent photoproduction
on the proton and on the nucleus have strongly different momentum transfer
distributions, which could  allow one to separate these contributions 
using a cut on the $J/\psi$ transverse momentum, $p_t$. 
While 
the contribution of coherent photoproduction on the nucleus can
dominate for small $p_t$ ($p_t \le 200$ MeV/c), 
it is strongly suppressed
by the nuclear form factor
for $p_t \ge 200$ MeV/c.

 Second, the ATLAS, CMS and ALICE detectors are equipped
with Zero Degree Calorimeters (ZDC) that can be effectively used to select events
of coherent photoproduction not accompanied by the strong interaction between
the proton and the nucleus. 
 Correspondingly, in theoretical estimates, one should
also suppress the strong interaction in the initial and final states. 
In the framework of the WW approximation, this is usually done by modifying the photon fluxes
emitted by protons and nuclei using the Glauber model of multiple
proton--nucleus scattering.

\section{Estimates of the photon fluxes in proton--nucleus UPCs}
\label{sec:fluxes_gz}

The expression for the photon flux of a fast moving non-point-like charged
particle with the charge $Z$ is given in many review papers (see, e.g., \cite{Bertulani:1987tz}):
\begin{equation}
N_{\gamma /Z}(\omega) \equiv \omega \frac{d N_{\gamma /Z}(\omega)}{d \omega}=
\frac{2Z^{2}\alpha_{{\rm em}}}{\pi}
\int_0^\infty dk_\bot
{k^{3}_\bot \left({F_{Z} (k^{2}_{\bot}+{{\omega ^2}/{\gamma ^{2}_{L}}})} 
\over {{k^{2}_{\bot}+{{\omega ^2}/{\gamma ^{2}_{L}}}}}\right)^2} \,,
\label{eq:pdflux} 
\end{equation}
where $\alpha_{{\rm em}}$ is the fine-structure constant; 
$F_{Z}(Q^2)$ is charge form factor of the particle ($F_{Z}(0)=1$) and
$\gamma _L $ is its Lorentz factor;
$\omega$ is the energy of the emitted photon.

In the proton case, one usually uses the dipole form of $F_{Z}(Q^2)$ in 
eq.~(\ref{eq:pdflux}), $F_{Z}(Q^2)=F_p(Q^2)=1/[1+Q^2/(0.71\,{\rm GeV}^2)]^2$.
With $F_{Z}(Q^2)=F_p(Q^2)$, the integral in eq.~(\ref{eq:pdflux}) can be readily calculated 
analytically or numerically. However, in proton--proton UPCs,
one frequently uses an approximate expression for $N_{\gamma /Z}(\omega)$~\cite{Drees:1988pp}
(see also~\cite{starlight}):
\begin{equation}
 N_{\gamma /p}(\omega) = {\frac {\alpha_{\rm em}} {2\pi}}
\left[1+\left(1-\frac {2 \omega} {\sqrt{s_{NN}}}\right)\right]
\left[\ln D-\frac {11} {6}+\frac {3} {D}-\frac {3} {2D^2}+\frac {1} {3D^3} \right] \,,
\label{DZflux}
\end{equation}  
where $D=1+0.71\,{\rm GeV^2}(\gamma_{L}^2/\omega^2)$.
Different approximations to the evaluation of $N_{\gamma /p}(\omega)$
are discussed in~\cite{Nystrand:2004vn}.

\begin{figure}[htb]
\centering
\epsfig{file=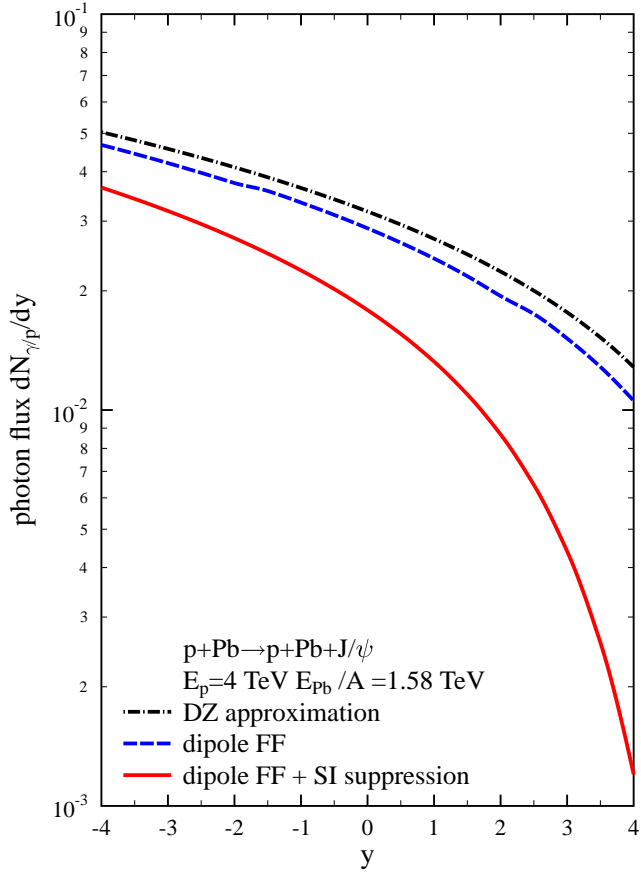,scale=0.5}
\vspace{-1 cm}
\caption{The flux of equivalent photons of the fast moving proton  
$N_{\gamma/p}(\omega)$ as a function of the $J/\psi$ rapidity $y$ 
in proton--Pb UPCs at $\sqrt{s_{NN}}=5.02$ TeV.
The curves are explained in text.}
\label{pPbflux}
\end{figure}

Figure~\ref{pPbflux} presents the flux of equivalent photons of the fast moving proton  
$N_{\gamma/p}(\omega)$ as a function of the rapidity $y$ of $J/\psi$ ($y = \ln (2\omega/M_{J/\psi})$) 
produced in proton--Pb UPCs at $\sqrt{s_{NN}}=5.02$ TeV.
In the figure, the exact calculation of eq.~(\ref{eq:pdflux}) is given by the blue dashed curve;
the approximate result of eq.~(\ref{DZflux}) is shown as the black dot-dashed curve
(labeled ``DZ approximation'').
One can see from the figure 
that while the two results agree well for large negative $y$ (corresponding to low $\omega$),
the difference between the results of eqs.~(\ref{eq:pdflux}) and (\ref{DZflux})
can reach up to 20\% for large $y$ in the region of the LHCb
measurement of proton--nucleus UPCs.

 In the case of proton--nucleus UPCs, one also needs to take into account the suppression of the 
strong interaction between colliding particles (see the discussion above). 
The resulting photon flux of the fast proton (nucleus) can be expressed as the following convolution over 
the impact parameter $b$ (the distance between the proton and nucleus centers of mass):
\begin{equation}
 N_{\gamma /Z}(\omega)= \int \limits_{0}^{\infty} d^2\vec{b} \, 
\Gamma_{pA}({\vec b}) \,N_{\gamma /Z}(\omega,\vec b) \,,
 \label{flux}
\end{equation}
where $N_{\gamma /Z}(\omega,\vec b)$ is the photon flux the transverse distance $\vec{b}$ away 
from the proton (nucleus) (see, e.g., \cite{vidovic}),
\begin{equation}
N_{\gamma /Z}(\omega,\vec b)=\frac{Z^2 \alpha_{\rm em}}{\pi^2}
\left(
\int_0^\infty dk_\bot
{{k^{2}_\bot F_{Z} (k^{2}_{\bot}+{{\omega ^2}/{\gamma ^{2}_{L}}})} 
\over {{k^{2}_{\bot}+{{\omega ^2}/{\gamma ^{2}_{L}}}}}}\, J_{1}(bk_{\bot})
\right)^2 \, ;
\label{bflux} 
\end{equation}
 $\Gamma_{pA}(\vec b)$ is the probability to suppress the proton--nucleus strong
interaction 
at small impact parameters $b$,
\begin{equation}
 \Gamma_{pA}({\vec b})=
 \exp\biggl (-\sigma_{NN}
 \int \limits^{\infty}_{-\infty}dz
 \rho_A(z,{\vec b})\biggr ) \,.
 \label{gamma}
\end{equation}
In eq.~(\ref{bflux}), $J_1$ is the Bessel function of the first kind.
In eq.~(\ref{gamma}), $\sigma_{NN}$  is the total nucleon--nucleon cross section 
at the corresponding $\sqrt{s_{NN}}$ (we use $\sigma_{NN} = 90$ mb at $\sqrt{s_{NN}} = 5.02$ TeV);  
$\rho_{A}(\vec r)$ is the nuclear density.

The photon flux $N_{\gamma /p}(\omega)$ calculated using eqs.~(\ref{flux})--(\ref{gamma}) is 
presented by the red solid curve in figure~\ref{pPbflux}. 
One can see from the figure that compared to the results of 
eqs.~(\ref{eq:pdflux}) and (\ref{DZflux}),
the strong proton--nucleus
interaction reduces the photon flux by the factor of $1.2 - 1.3$ at low photon energies 
(large negative $y$), 
by the factor of two at central rapidities, and strongly suppresses  $N_{\gamma /p}(\omega)$
at large 
rapidities~\footnote{  
The presented result does not include the effect that the proton interacts
not with the entire nucleus located at its center, but with the nucleons inside the nucleus 
whose
spatial distribution is given by the nuclear density. The inclusion of this effect
leads to some increase of  $N_{\gamma /p}(\omega)$ at large $\omega$~\cite{Baron:1993nk}.}.

\begin{figure}[t]
\centering
\epsfig{file=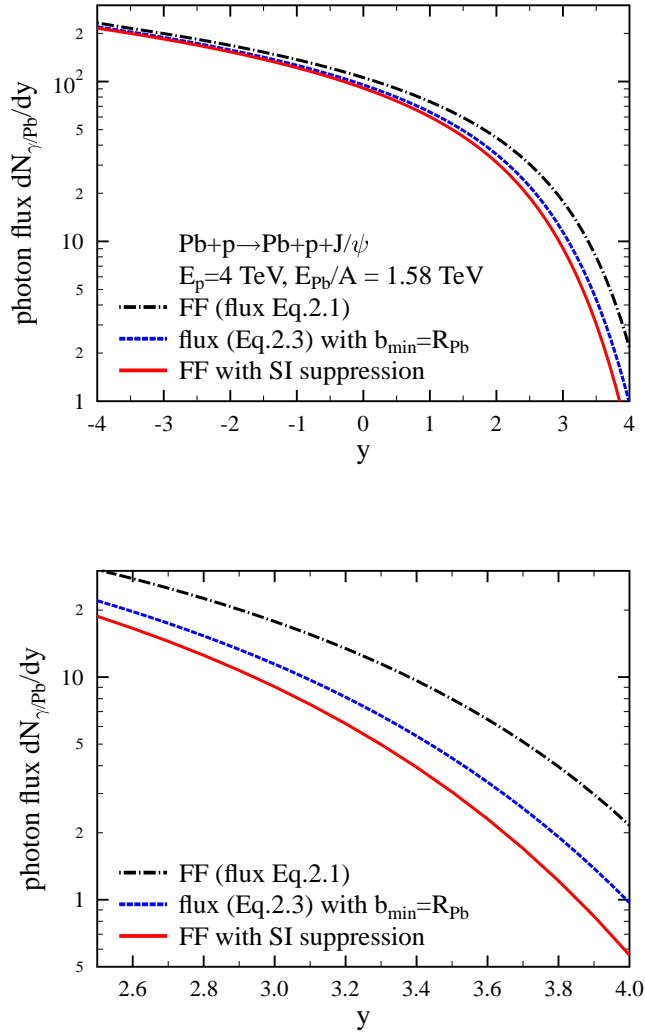,scale=0.5}
\caption{The flux of equivalent photons of a fast moving Pb nucleus,   
$N_{\gamma/Pb}(\omega)$, as a function of the $J/\psi$ rapidity $y$ 
in Pb--proton UPCs at $\sqrt{s_{NN}}=5.02$ TeV. The curves are explained in text.
Both panels present the same information with 
the lower panel highlighting the region of large rapidities of $2.5 \leq y \leq 4$.
}
\label{Pbpflux}
\end{figure}

The photon flux generated by a fast moving nucleus can be calculated
using eqs.~(\ref{flux})--(\ref{gamma}) with the appropriate nuclear charge form factor. 
In the case of  Pb, we used the nuclear density distribution obtained
in the Hartree--Fock--Skyrme model,
which describes well the root-mean-square charge radius of Pb and elastic
electron--Pb scattering.  The resulting
photon flux of a fast moving Pb nucleus,  
$N_{\gamma/Pb}(\omega)$, as a function of the rapidity $y$ of $J/\psi$ produced in 
Pb--proton UPCs at $\sqrt{s_{NN}}=5.02$ TeV is presented by the red curve in figure~\ref{Pbpflux}.

For comparison, in figure~\ref{Pbpflux} we also present $N_{\gamma/Pb}(\omega)$ calculated 
using the following two approximations. First, the blue dashed curve corresponds to 
the calculation neglecting the suppression of the 
strong
proton--nucleus interaction by setting
 $\Gamma_{pA}({\vec b})=1$ in eq.~(\ref{flux}) and using instead the lower limit on the integration over
$|\vec{b}|$, $b_{\rm min}=R_{Pb}$ ($R_{Pb}$ is the radius of Pb). 
Second, the black dot-dashed curve corresponds to the evaluation of
 $N_{\gamma/Pb}(\omega)$ using eq.~(\ref{eq:pdflux}) with the Pb charge form factor.

Note that while both panels in figure~\ref{Pbpflux} present the same information, the lower one
highlights the region of large rapidities of $2.5 \leq y \leq 4$ important for the 
measurement of 
coherent  $J/\psi$ production by high-energy photons in  
Pb--proton UPCs at the LHC.

To simplify calculations, one frequently uses the following approximate analytical expression 
for the photon flux of a fast moving nucleus: 
\begin{equation}
N_{\gamma /Z}(\omega)=\frac {2Z^{2}\alpha_{\rm em}} {\pi} \left[\zeta K_{0}(\zeta )K_{1}(\zeta)-
{\frac {{\zeta}^2} {2}(K_{1}^2 \left(\zeta ) - K_{0}^2 (\zeta)\right)}\right] \,,
\label{plflux}
\end{equation}
where $K_{0}$ and $K_{1}$ are the modified Bessel functions of the second kind; 
 $\zeta =\omega b_{\rm min}/\gamma_L$, where $b_{\rm min}$ is the minimal
admitted distance in the impact parameter space chosen to suppress the strong interaction between
the colliding particles. In the considered case of the proton--Pb interaction, 
it is reasonable to take $b_{\rm min} \approx (1.1 - 1.2)R_{Pb}$.
In particular, with 3\% accuracy, eq.~(\ref{plflux}) with  $b_{\rm min}=1.15R_{Pb}$ 
reproduces the exact result of eqs.~(\ref{flux})--(\ref{gamma}) (the red solid curve in figure~\ref {Pbpflux})
and with $b_{\rm min}=R_{Pb}$ --- the result presented by the blue dashed curve.

\section{Cross section of coherent $J/\psi$ 
photoproduction in LO pQCD}
\label{sec:cs_gz}

Equation~(\ref{csupc}) allows one to calculate the rapidity distribution  
of  $J/\psi$ photoproduction in proton--nucleus UPCs.
In obtaining the results presented below, we used the exact results for the photon flux
of the proton, $N_{\gamma/p}(\omega)$ [eq.~(\ref{eq:pdflux}) and the red solid curve in figure~\ref{pPbflux}] 
and for the photon flux of Pb, $N_{\gamma/Pb}(\omega)$ [eqs.~(\ref{flux})-(\ref{gamma}) and the red solid 
curve in figure~\ref{Pbpflux}] and the results of our leading order (LO) pQCD analysis of exclusive 
$J/\psi$ photoproduction on the proton and nuclei~\cite{Guzey:2013qza}.

To recapitulate main results of~\cite{Guzey:2013qza}, at the leading order, the cross section
of exclusive $J/\psi$ photoproduction on the proton reads:
 \begin{eqnarray}
\sigma_{\gamma p\to J/\psi p}(W_{\gamma p})=
\frac{M^3_{J/\psi}\Gamma_{ee}{\pi}^3}{48\,\alpha_{\rm e.m.}\mu^8} \frac{1}{B_{J/\psi}(W_{\gamma p})}
(1+\eta^2)\,F^2(\mu^2)
\left[R_g \alpha_s(\mu^2) xG_p(x,\mu^2 )\right]^2 \,,
\label{csprot}
\end{eqnarray}
where $\Gamma_{ee}$ is the width of the $J/\psi$
electronic decay; $B_{J/\psi}(W_{\gamma p})$ is the slope of the $t$ dependence of the
$\gamma p\to J/\psi p$ cross section; $\alpha_s(\mu^2)$ is the 
strong running coupling constant; 
$G_p(x,\mu^2)$ is the gluon density of the proton;  $x=M_{J/\psi}^2/W_{\gamma p}^2$;
$\eta$ is the ratio of the real to the imaginary parts of the $\gamma p\to J/\psi p$
amplitude; $R_g$ is the enhancement factor taking into account the effect of skewness 
in the exclusive  $\gamma p\to J/\psi p$ reaction;
 $F^2(\mu^2)$ is the suppression factor stemming from a host of effects beyond the approximation used
in eq.~(\ref{csprot}) (next-to-leading order corrections, the effect of the overlap
between the photon and $J/\psi$ wave functions, etc.).

The analysis of~\cite{Guzey:2013qza} demonstrated that the HERA and LHCb data on
$J/\psi$ photoproduction on the proton can be described very well by eq.~(\ref{csprot}) evaluated at 
 the hard scale of $\mu^2=3$ GeV$^2$ 
using
a large array of modern gluon distributions
in the proton.
The results presented below (including the nuclear case)
 are calculated with the MNRT07 gluon distribution~\cite{Martin:2007sb}: 
since this distribution is constrained to describe the HERA $J/\psi$ photoproduction data, 
the corresponding suppression factor is absent, i.e., $F^2(\mu^2)=1$.

When extending eq.~(\ref{csprot}) to the case of a nuclear target, one needs to take into
account the effect of the leading twist nuclear gluon shadowing~\cite{Frankfurt:2011cs} and 
the fact that $\eta$ and 
$R_g$ for the $\gamma A\to J/\psi A$ amplitude are smaller than their proton 
counterparts~\footnote{At small $x$, the leading twist nuclear gluon shadowing slows down an 
increase of the nuclear gluon density with a decrease of $x$, which leads to a decrease of 
$\eta$ and $R_g$.}; the combination of these two effects is encoded in the nuclear suppression
factor $S_A(W_{\gamma p})$~\cite{Guzey:2013xba,Guzey:2013qza}. 
 The resulting cross section
of exclusive $J/\psi$ photoproduction on a nucleus is:
\begin{equation}
\sigma_{\gamma A \rightarrow J/\psi A}(W_{\gamma p})=
S_{Pb}^2(W_{\gamma p})\,\frac{d\sigma_{\gamma p\rightarrow J/\psi p}(W_{\gamma p},t=0)}{dt} \,
{\Phi_A(t_{\rm min})} \,,
\label{csa1} 
\end{equation} 
where
$\Phi_A(t_{\rm min})=\int \limits_{t_{\rm min}}^{\infty} dt \left|F_A (t)\right|^2$,
where $t_{\rm min}=-M_{J/\psi}^4 m_N^2/W_{\gamma p}^4$ is the minimal momentum transfer to the nucleus and
$F_A (t)$ is the nucleus form factor.

\section{Results and discussion}
\label{sec:results_gz}

\begin{figure}[h]
\centering
\epsfig{file=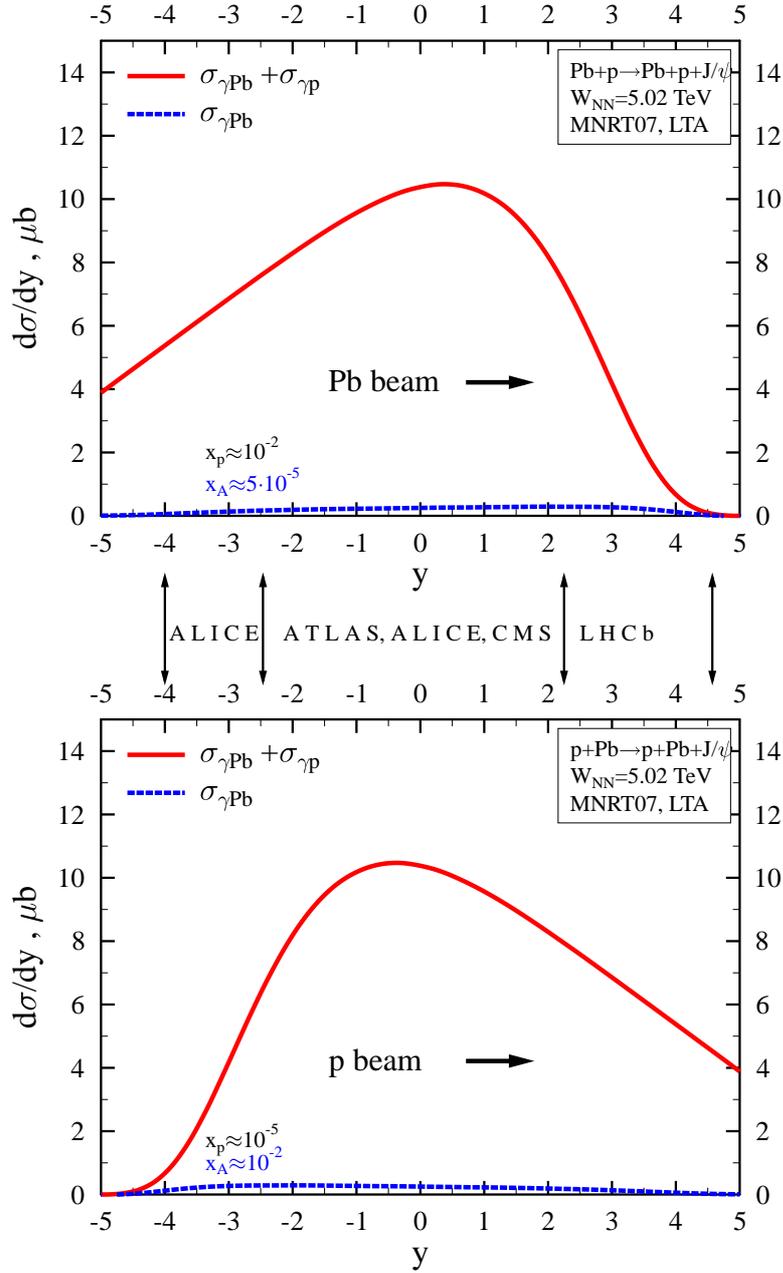,scale=0.6}
\caption{The $t$-integrated rapidity distribution of $J/\psi$ photoproduction in lead--proton 
(upper panel) and proton-lead (lower panel) UPCs
at $\sqrt{s_{NN}}=5.02$ TeV.}
\label{rapidpbp}
\end{figure}

Figure~\ref{rapidpbp} presents our predictions for the rapidity distribution 
of $J/\psi$ photoproduction in proton--Pb UPCs integrated over 
the momentum transfer $t$  in the LHC kinematics 
calculated using eq.~(\ref{csupc}) and the input discussed above.
In the figure, the red solid curves correspond to the sum of both terms in eq.~(\ref{csupc});
the blue dashed curves are the contribution of the photon--nucleus term only.
One expects that the photon--proton contribution should by far dominate the photon--nucleus
one because (i) the nuclear enhancement of the photon flux emitted by the nucleus (the factor
of $Z^2$) is much stronger than that of the $t$-integrated coherent photon-nucleus cross section
(the factor of $A^{4/3}$) and (ii) because nuclear shadowing suppresses the photon--nucleus cross 
section.
One can
readily see from the figure that the contribution of
photoproduction of $J/\psi$ on the proton  dominates in the whole range
of rapidities which can be studied by the
ALICE, ATLAS, CMS and LHCb detectors 
(the ranges of $y$  covered by
the corresponding experiments are indicated by the 
labels ``ALICE'', ``ATLAS, ALICE, CMS'' and ``LHCb'').

Since during the $pA$ run in 2013 the beam direction was inverted,
we show two options of the collision geometry.
 In the top panel of figure~\ref{rapidpbp}, the 
$J/\psi$ rapidity 
$y$ is positive in the nucleus beam direction. 
In this case, the ALICE muon detector covering the rapidity range  
of $-4<y<-2.5$  
probes (i)
production of $J/\psi$ on the proton by low-energy photons
emitted by lead 
(the proton gluon density around $x_p \approx 10^{-2}$ for $y \approx -3$)  
and (ii) production $J/\psi$ on the nucleus
by high-energy photons emitted by the proton 
(the nuclear gluon distribution down to $x_A \approx 10^{-5}$).
The corresponding average values of the probed $x$ are indicated in the figure.

One should note that with the considered beam directions, 
the LHCb detector, 
which covers the $2 < y <4.5$ range, 
can measure photoproduction on the proton by 
high-energy photons emitted by lead and, hence, can access 
the gluon distribution in the proton down to $x_p\approx 10^{-5}$.

The bottom panel of  figure~\ref{rapidpbp} corresponds to the inverse beam direction. 
Therefore, ALICE will study $J/\psi$ photoproduction in 
the interaction of high-energy photons with the proton
(small $x_p$) and of low-energy photons with Pb (around $x_A \approx 10^{-2}$); 
the LHCb detector will access scattering of low-energy photons on the proton and
of high-energy photons on Pb.

From the calculated rapidity distributions 
presented in figure~\ref{rapidpbp}, 
we find that the contribution 
of $J/\psi$ photoproduction on the nucleus is small -- its contribution 
ranges from 2\% for the high-energy 
photon--nucleus photoproduction to about 7\%  for the low-energy photons.
Note that since our theoretical description~\cite{Guzey:2013xba} of the 
$\gamma  Pb\rightarrow J/\psi Pb$ cross section at  
$x_A \approx 10^{-2}$ and $x_A \approx 10^{-3}$ reproduces well the ALICE data~\cite{alice1,alice2}
and very weakly depends on the choice of the gluon distribution and the hard scale $\mu^2$,
the proton--nucleus contribution can be considered to be reliably fixed.
Therefore,
one can try to exclude it using a cut on the momentum 
transfer $t$ since the momentum transfer distributions in photoproduction
on the nuclear and the proton targets are strongly different.  

\begin{figure}[htb!]
\centering
\epsfig{file=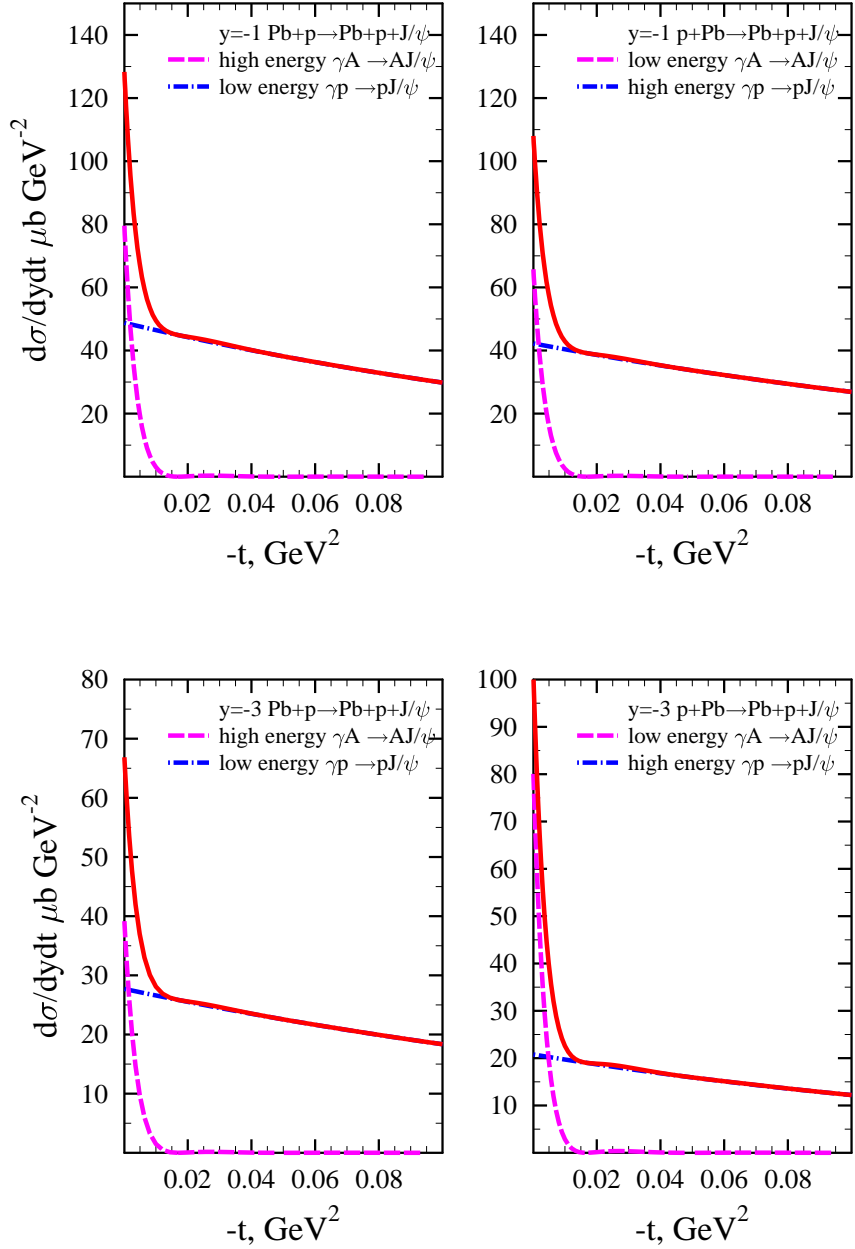,width=12cm}
\caption{
The distribution 
of coherent $J/\psi$ photoproduction in Pb-proton (left panels) and proton-Pb (right panels)
UPCs
as a function of $t$ for $y=-1$ and $y=-3$ 
at $\sqrt{s_{NN}}=5.02$ TeV.}
\label{fig:fig5}
\end{figure}

To check whether it is possible
to separate coherent photoproduction of $J/\psi$ on the nuclear and proton targets
in the current kinematics,  
we calculated the distribution  
of coherent $J/\psi$ photoproduction in proton-Pb and Pb-proton UPCs in the LHC kinematics
as a function of the  momentum transfer squared $t$ at a few values 
of the rapidity $y$ (see figure~\ref{fig:fig5}). 
As expected, the contribution of photoproduction 
on the nuclear target is strongly peaked at very small $-t$. Thus, it  can be   
either separated
by the cut on small $t$, $|-t|\le 0.02$ GeV$^2$, 
or analyzed provided the experiment has  sufficiently high statistics.

\begin{figure}[htb!]
\centering
\epsfig{file=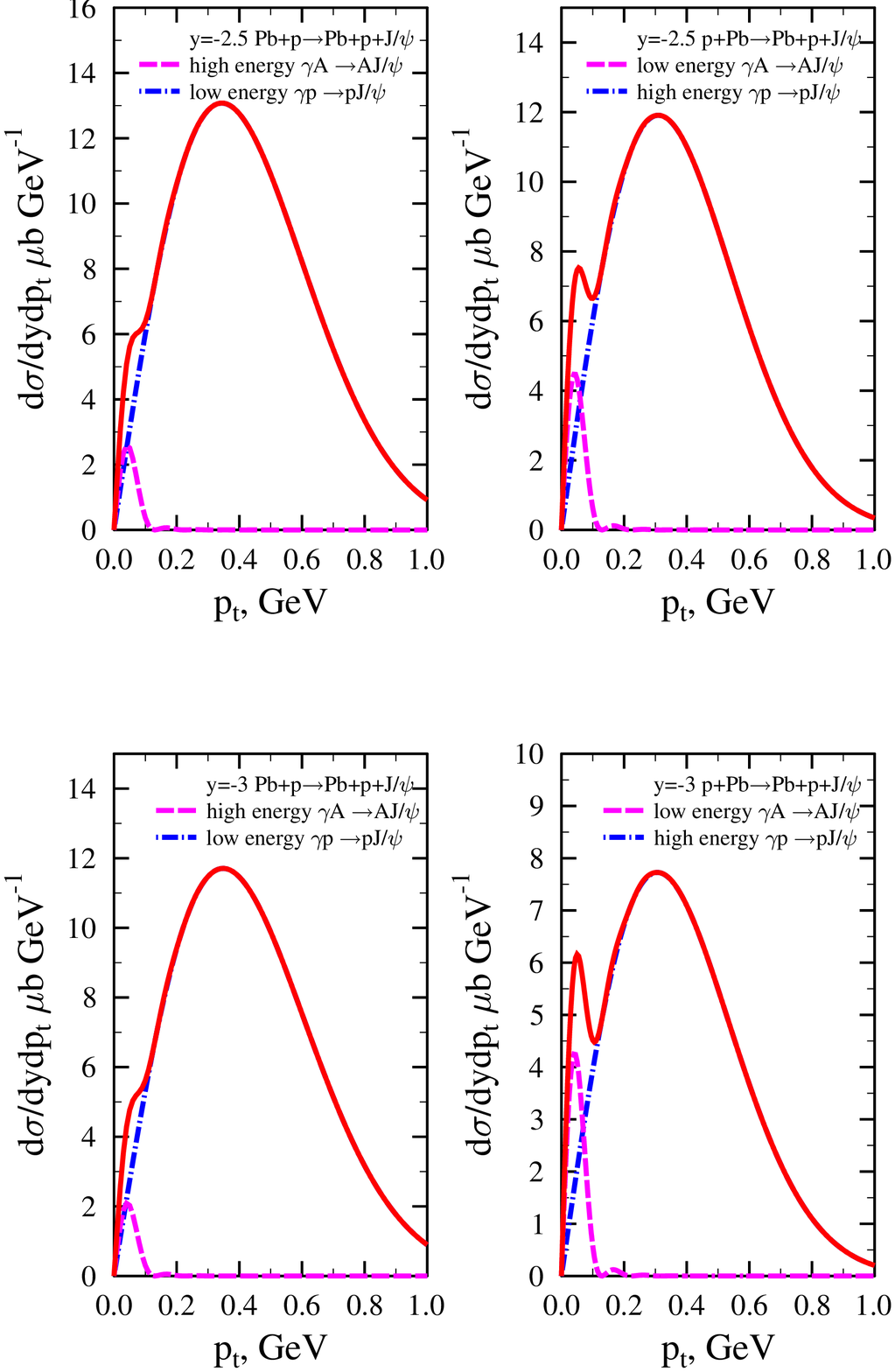,width=12cm}
\caption{The distribution 
of coherent $J/\psi$ photoproduction in Pb-proton (left panels) and proton-Pb (right panels)
UPCs
as a function of the momentum transfer $p_t$ for $y=-2.5$ (upper panels) and $y=-3$ (lower panels)  
at $\sqrt{s_{NN}}=5.02$ TeV.
}
\label{fig:fig6}
\end{figure}

\begin{figure}[htb!]
\centering
\epsfig{file=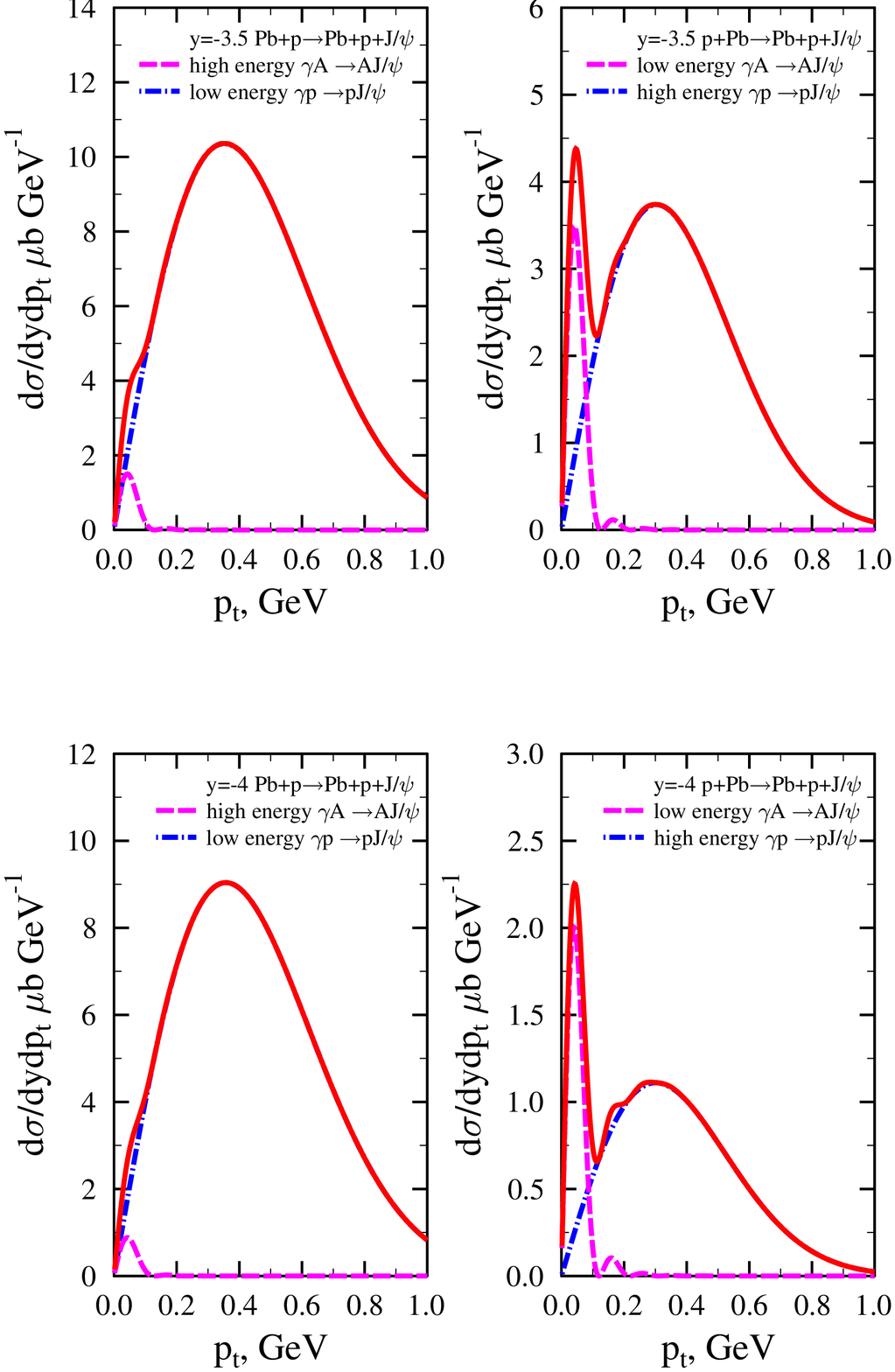,width=12cm}
\caption{The same as in figure~\ref{fig:fig6}, but for  $y=-3.5$ (upper panels) and $y=-4$ (lower panels).}
\label{fig:fig7}
\end{figure}

There is an interesting feature which can be seen when the distribution of 
coherent $J/\psi$ photoproduction in proton-Pb and Pb-proton UPCs is presented
as a function of
the transverse momentum transfer $p_t$ (see figures~\ref{fig:fig6} and \ref{fig:fig7}).
Coherent photoproduction on lead by low-energy photons from the proton
results in a narrow peak in the momentum transfer distribution starting
from the rapidities of  $y \le -2.5$ (the same effect can be observed at
positive rapidities when the  beam direction is reversed).  
For the rest of rapidities ($y > -2.5$), photoproduction on the proton target 
is significantly larger and, thus, this peak disappears.
This effect arises due to the steep drop of the photon flux 
generated by Pb with an increase of the photon energy.   
While experimentally this effect can be revealed
only with the high transverse momentum resolution, 
this should 
nevertheless
result in some
excess of events at small $p_t<150$ MeV/c . 
It is important to follow our suggestions for the cuts in $t$ and $p_t$ by detailed numerical 
studies/simulations examining how these cuts can be realized in the LHC experiments.
 
It is of interest to check feasibility to quantify the magnitude of the
nuclear gluon shadowing
and, thus, to differentiate among different theoretical predictions for nuclear shadowing, 
 using the analysis of the shape of transverse
momentum distributions  measured in 
coherent $J/\psi$ photoproduction in 
ultraperipheral proton-Pb 
collisions at $\sqrt{s_{NN}} =5.02 \,{\rm TeV}$. 

From the ALICE measurements of coherent photoproduction in Pb-Pb UPCs at
$\sqrt{s_{NN}}=2.76$ TeV, it was found that at $x \approx 10^{-3}$, the nuclear
gluon shadowing is $R_g(x \approx 10^{-3}) \approx 0.6$~\cite{Guzey:2013xba}. 
This
value is in a good agreement with 
the predictions of the EPS09LO fit and 
the leading twist approximation (LTA) in the theory of nuclear shadowing employing a large array 
(CTEQ6L, CTEQ6L1, MRST04, NNPDF, and MNRT07LO) of 
leading order gluon distributions in the proton, for details and references, see~\cite{Guzey:2013qza}.
While the ALICE Pb-Pb UPC data does not allow one discriminate between the EPS09LO and the LTA 
approaches since their respective predictions for $R_g$ converge at $x \approx 10^{-3}$,  
in the current proton--Pb study, the ALICE and LHCb muon spectrometers
extend the kinematic coverage in $x$ down to  $x\approx 5 \times 10^{-5}$, where the difference
between the EPS09LO and LTA+MNRT07LO predictions for $R_g$ is sizable. 
Indeed, while EPS09LO predicts that $R_g$ is practically constant for $x < 10^{-3}$ with
$R_g \approx 0.6$, 
the LTA+MNRT07LO nuclear gluon shadowing increases with a decrease of $x$ and reaches
 $R_g(x\approx 5\times 10^{-5}) \approx 0.4$. 
This difference in the predicted values of $R_g(x\approx 5\times 10^{-5})$ leads 
to the approximately factor of two difference in the predicted values of the 
$\gamma A \to J/\psi A$ cross section.  
In spite of the smallness of the $\gamma A \to J/\psi A$ contribution to the $p+Pb \to p+Pb+J/\psi$ 
process (see figures~\ref{fig:fig6} and \ref{fig:fig7}),
this difference can be seen in the transverse momentum distribution.

\begin{figure}[t]
\centering
\epsfig{file=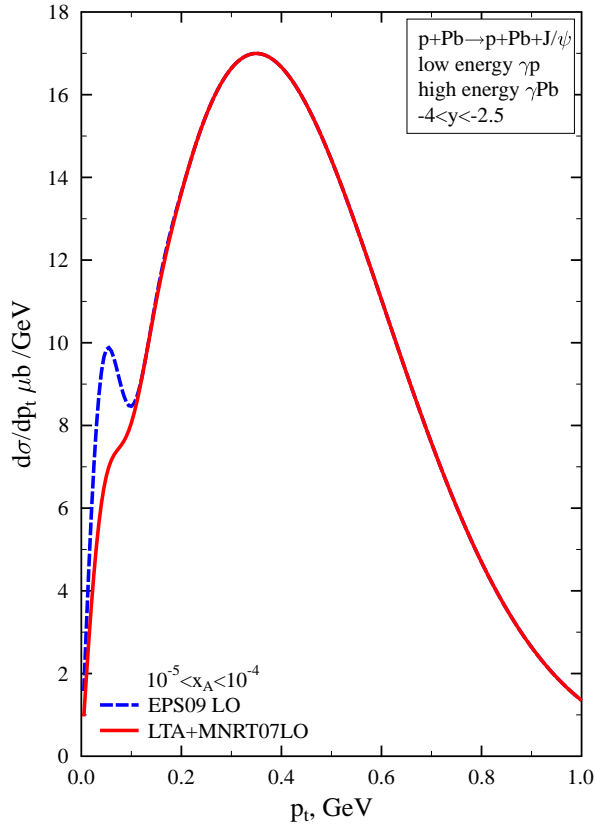,scale=0.47}
\vspace{-1 cm}
\caption{
The $p+Pb \to p+Pb+J/\psi$ 
transverse momentum distribution 
integrated over the $-4.0<y<-2.5$ range at $\sqrt{s_{NN}}=5.02$ TeV. See text for details.}
\label{fig:diff}
\end{figure}

Figure~\ref{fig:diff} presents the result of our calculations of
the $p+Pb \to p+Pb+J/\psi$ 
transverse momentum distribution 
integrated over the $-4.0<y<-2.5$ rapidity interval. Note that 
for the ALICE detector, the positive 
rapidity $y$ is in the Pb beam direction. (In the case of the LHCb detector,
this kinematics corresponds to $2.5 <y <4$ in pPb collisions with the positive rapidity in 
the direction of the proton beam.)  
The red solid curve corresponds to the LTA+MNRT07LO prediction for the nuclear gluon shadowing
(see the left panels in figures~\ref{fig:fig6} and \ref{fig:fig7}); 
the blue dashed curve corresponds to 
the central value of the EPS09LO fit.
As we explain below, we do not show the small error band around each curve since its effect
is negligibly small.

One can see from figure~\ref{fig:diff} that the two scenarios of the nuclear gluon shadowing
predict different shapes of the transverse momentum distribution at small $p_t$.
In particular, since the small-$x$ gluon shadowing is weaker in EPS09LO than in LTA+MNRT07LO,
 a peak---about 30\% excess---appears at small $p_t$ in the EPS09LO case. 
Since the shape of the momentum transfer 
distribution in $\gamma +p\rightarrow J/\psi+p$ in this region of 
$W_{\gamma p}$ is well known ($d\sigma /dt \propto \exp(B t)$),
an observation or non-observation of a small-$p_t$ shoulder in the 
$p+Pb \to p+Pb+J/\psi$ transverse momentum distribution would 
be unambiguously correlated with the magnitude of the  
nuclear gluon shadowing. Thus, such a measurement could be
the first experimental estimate of the
nuclear gluon shadowing at $x\approx 5\times 10^{-5}$ in lead.

Note that the analysis and interpretation of $p+Pb \to p+Pb+J/\psi$ data require
taking into account the following three effects. First,
the $\gamma \gamma \rightarrow \mu \mu$ process
contributing at small $p_t$ should be subtracted. 
The cross section of this process is
reliably calculated in StarLight.
Second, the $\gamma +p\rightarrow J/\psi +X$ process with 
diffractive dissociation of the proton target distorts the shape
of the transverse momentum distribution.
However, since this process was
studied  in this region of $x$ at HERA,
its contribution can be easily modeled.
Third, the target dissociation in the $\gamma A$
process can be rejected by a ZDC.

It should be emphasized that the numerical predictions using eqs.~(\ref{csprot}) and (\ref{csa1}), 
which we show in this section, employ the MNRT07 gluon density at $\mu=3$ GeV$^2$.
This parameterization describes very well the data on $J/\psi$ photoproduction on the proton 
and well the data on coherent $J/\psi$ photoproduction on Pb~\cite{Guzey:2013qza}.
A similarly good description of these data sets can be obtained with other
parametrizations of the gluon distributions of the proton (CTEQ6L, CTEQ6L1, MRST04, NNPDF) 
and the corresponding nuclear suppression factors $S_A(W_{\gamma p})$, which could be evaluated 
at a range of $\mu^2$ near $\mu=3$ GeV$^2$, $\mu=2.4-3.4$ GeV$^2$. 
Thus, different choices of the gluon distribution and the scale $\mu^2$ do not affect
our conclusions. 
At the same time, the use of a different gluon parameterization evaluated at a slightly different
scale $\mu^2$ will affect our predictions for the photon--nucleus cross section at high 
energies corresponding to $x_A \ll 10^{-3}$, see the corresponding curves in figs.~\ref{fig:fig5},
\ref{fig:fig6} and \ref{fig:fig7}. An example of sensitivity to this effect is presented 
in fig.~\ref{fig:diff}.

Note also that we do not show the theoretical uncertainty of the photon--nucleus contribution, 
which comes from the uncertainties of the predicted amount of nuclear gluon 
shadowing in the framework of the leading twist approximation (LTA+MNRT07L0). 
Its effect is small compared to the magnitude and 
pattern of the rapidity distributions presented in this section.

Our results can be compared to the predictions for $J/\psi$ photoproduction in
proton--nucleus UPCs available in the literature.
The approach used in~\cite{Adeluyi:2013tuu} is very similar to ours but different in implementation.
First, the photon flux of the proton 
 used in~\cite{Adeluyi:2013tuu} does not include 
the effect of the suppression of the proton--nucleus strong interaction at small impact parameters. 
As a result, the maximum of the rapidity distribution of 
$J/\psi$ photoproduction on the nucleus is shifted to 
significantly higher $W_{\gamma p}$.
Second, the MSTW08 gluon distribution~\cite{Martin:2009iq} used in~\cite{Adeluyi:2013tuu} 
fails to describe the LHCb data on the $W_{\gamma p}$ behavior of the $\gamma p \to J/\psi$ cross 
section. This gluon distribution leads to a strong increase of the cross
section at high photon energies (small gluon $x$) resulting in a
significant shift (by two units of rapidity) of the maximum of the rapidity distribution of 
$J/\psi$ photoproduction on the proton in Pb--proton UPCs compared to our results
in figure~\ref{fig:fig5}. 
As a result, in the range of rapidities $y$ corresponding to high-energy photons emitted by Pb, 
our predictions differ by as much as a factor of three.   
Third, the combination of the MSTW08 gluon distribution with the nuclear PDFs 
extracted from the global QCD fits (such as the EPS09 nuclear PDFs~\cite{eps09}) is inconsistent, 
see the discussion in~\cite{Guzey:2013qza}.

Predictions for the rapidity distribution of the 
$Pb+p \to Pb+p+ J/\psi$ cross section were also made 
using the framework of the color dipole model~\cite{Lappi:2013am}. The resulting distribution 
is quantitatively similar to our result. However, one has to keep in mind that the dipole approach
overestimates the $Pb Pb \to Pb Pb J/\psi$ cross section measured by the ALICE 
collaboration~\cite{alice1},
see the discussion in~\cite{Guzey:2013qza}.

Note also that the momentum transfer distributions have not been analyzed in 
\cite{Adeluyi:2013tuu} and \cite{Lappi:2013am}.

\section{Conclusions}
\label{sec:conclusions_gz}

In conclusion, we have shown that the study of $J/\psi$ photoproduction in 
Pb-proton and proton-Pb UPCs
at the LHC energies allows one to measure with good accuracy photoproduction of 
charmonium on the proton target at small $x$, when 
one imposes a 
cut on the transverse momentum of produced $J/\psi$ at the
level of $p_{t}\ge 150$ MeV/c. 
It will be  hardly possible to extract 
the cross section of photoproduction on a nucleus at small $x$ and, hence, to 
quantify the effect of the nuclear gluon shadowing since even after
applying the $p_{t}\le 150$ MeV/c cut, the $\gamma p \to J/\psi p$ contribution
at large $x$ is still comparable to the 
$\gamma A \to J/\psi A$ 
contribution at small $x$.
  By analyzing the momentum transfer distributions, we found that
an interplay of  $J/\psi$ production by low-energy photons on the nucleus
and by high-energy photons on the proton in proton--Pb UPCs 
can result in some excess of events at small $p_t$ in a definite 
region of rapidities (for $y < -2.5$). Such an excess can be 
studied
by the ALICE and LHCb collaborations.

\section*{Acknowledgements}

We would like to thank E.~Kryshen for useful discussions.

\end{document}